\documentstyle[12pt]{article}
\begin{document}

\newcommand{\be}{\begin{equation}}
\newcommand{\ee}{\end{equation}}

\baselineskip 16pt

\begin{flushright}
EFI-02-75 \\ hep-th/0204247 \end{flushright}

\vspace*{2cm}
\begin{center}
{\Large{\bf{Localized Tachyons and}}}\\

{\Large{\bf{the $g_{cl}$ conjecture}}}

\vspace*{0.3in}
Anirban Basu
\vspace*{0.3in}

\it
Enrico Fermi Institute and Department of Physics,
University of Chicago\\
5640 S.~Ellis Avenue, Chicago, IL~60637, USA\\
{\tt basu@theory.uchicago.edu} \\
\vspace*{0.2in}

\end{center}
\begin{abstract}
We consider ${\mathcal{C}} / {\mathcal{Z}}_N$ and ${{\mathcal{C}}^2} / {\mathcal{Z}}_N$ orbifolds of heterotic string theories and $\mathcal{Z_{N}}$ orbifolds of $AdS_3$. We study theories with $\mathcal{N}$=2 worldsheet superconformal invariance and construct RG flows. Following Harvey, Kutasov, Martinec and Moore, we compute $g_{cl}$ and show that it decreases monotonically along RG flows- as conjectured by them. For the heterotic string theories, the gauge degrees of freedom do not contribute to the computation of $g_{cl}$.
\end{abstract}
\vfill

\newpage

\section{ Introduction}

In the recent past, the condensation of closed string tachyons has been the subject of active research. In a paper by Adams, Polchinski and Silverstein (APS) [1], the authors considered closed string tachyons which are localized at the fixed points of noncompact nonsupersymmetric orbifolds of string theory [2,3,4,5]. They studied the fate of the condensation of the tachyon and argued that the condensation smoothens out the the tip of the orbifold singularity, leading to flat space or ALE space. Thus closed string tachyon condensation is seen to be associated with the decay of spactime itself. In a paper by Harvey, Kutasov, Martinec and Moore (HKMM) [6] the problem of closed string tachyon condensation was analyzed further. They used world sheet renormalization group techniques to study the condensation of tachyons. In particular, they considered a class of theories with (2,2) worldsheet superconformal invariance and studied RG flows which preserve this invariance. They defined a quantity called ``$g_{cl}$'', and they conjectured that it should decrease monotonically along RG flows. They computed $g_{cl}$ for the theories they had considered and showed that $g_{cl}$ is indeed a monotonically decreasing function along RG flows from the UV to the IR. Their analysis also reaches the same conclusion as the APS analysis, namely, that tachyon condensation smoothens out the tip of the singularity. However, some of the flows which are expected to occur according to the APS analysis are forbidden to occur by the HKMM analysis, as $g_{cl}$ does not decrease along these flows.\\
 In this paper, we will test the $g_{cl}$ conjecture put forward by HKMM for noncompact orbifolds of heterotic string theory [7,8] and ${\mathcal{Z}}_N$ orbifold of $AdS_3$ [9,10]. We will see that the conjecture holds and that $g_{cl}$ is monotonically decreasing along RG flows. We shall also see that the results for RG flows, where they exist, are similar to those obtained for type 0 and type II theories by HKMM. In particular, in such cases the gauge degrees of freedom in heterotic string theory do not play a significant role in tachyon condensation, apart from the level-matching constraints. Because of the similarity of the results with the HKMM analysis, we will be brief and quote the results without giving much details, for which one can look at the HKMM paper.\\
In section 2, we construct the constraints from  modular invariance in heterotic string theories. In section 3, we analyse ${\mathcal{C}} / {\mathcal{Z}}_N$ flows in heterotic string theories and in section 4 we analyse ${\mathcal{C}}^2/{\mathcal{Z}}_N$ flows. In section 5, we consider RG flows for the  ${\mathcal{Z}}_N$ orbifold of $AdS_3$. Finally, in section 6 we present the conclusions.\\

\section{Constraints from Modular invariance}
We shall be considering three heterotic string theories in this paper- the nonsupersymmetric $SO(32)$ heterotic string theory [11,12], and the supersymmetric $SO(32)$ and $E_8 \times E_8$ heterotic string theories. We shall be studying ${\mathcal{C}}/{\mathcal{Z}}_N$ and ${\mathcal{C}}^2/{\mathcal{Z}}_N$ orbifolds of these theories. As is well known there are nontrivial constraints on the action of the orbifold group arising from modular invariance in the heterotic string theories [13,3]. The constraints arising in the various theories are as follows:\\
\subsection{$E_8 \times E_8$}
 We shall be working in the fermionic formulation of the heterotic string theory. We choose a complex basis $Z_j$ where $Z_j=X_{2j}+i X_{2j+1}$ for $j$=1,2,3,4 for the world sheet bosons and similarly for ${\overline{\psi}}_j$, the right-moving world sheet fermions. Similarly we choose a complex basis $\lambda_a$ for $a$=1,...,16 corresponding to the 32 Majorana-Weyl fermions in the left-moving CFT. We diagonalize the action of the orbifold group G on these degrees of freedom. The orbifold action is\\
\be
 G: (Z_1,Z_2,Z_3,Z_4) \rightarrow (e^{\frac{2\pi i r_{1}}{N}} Z_1,e^{\frac{2\pi i r_{2}}{N}} Z_2,e^{\frac{2\pi i r_{3}}{N}} Z_3, e^{\frac{2\pi i r_{4}}{N}} Z_4)\
\label{feq1}
\ee
 and similarly for the right moving fermions. Also the action of G on the gauge degrees of freedom is given by\\
\be
 G:(\lambda_1,\lambda_2,...,\lambda_{8}) \rightarrow (e^{\frac{2\pi i p_1}{N}}\lambda_1,e^{\frac{2\pi i p_2}{N}}\lambda_2,...,e^{\frac{2\pi i p_{8}}{N}}\lambda_{8})\
\label{feq2}
\ee
\be
G:(\lambda_9,...,\lambda_{16}) \rightarrow (e^{\frac{2\pi i q_1}{N}} \lambda_9,e^{\frac{2\pi i q_2}{N}}\lambda_{10},...,e^{\frac{2\pi i q_8}{N}}\lambda_{16})\
\label{feq3}
\ee
 Because we have fermions in the theory, we have to define the theory on the eight fold cover of $SO(9,1) \times SO(16) \times SO(16)$, i.e. Spin$(9,1) \times$ Spin$(16) \times$ Spin$(16)$, and hence on demanding that we have an orbifold of order $N$, we get the constraint that for $N$ even\\
\be
\sum_{i=1}^{4} r_i = \sum_{a=1}^{8} p_a = \sum_{a=1}^{8} q_a=0	\, \rm{mod} \, 2\
\ee
For $N$ odd, this does not lead to an independent constraint. The constraints from  modular invariance are\\
\be
 \sum_{i=1}^{4} r_i ^2 - \sum_{a=1}^{8} p_a ^2 - \sum_{a=1}^{8} q_a ^2=0 \, \rm{mod}\, {\it{N}} \, \rm{for} \, {\it{N}} \, \rm{odd}
\ee
\be
 {\hspace*{1.8in}}=0 \, \rm{mod} \, \, 2{\it{N}} \, \rm{for} \,  {\it{N}} \, \rm{even}.\
\ee

\subsection{Non-supersymmetric $SO(32)$ and supersymmetric $SO(32)$}\
  The action of $G$ on the $Z_i$'s and the right moving fermions remain as before while the action on the left moving gauge degrees of freedom is\\
\be
G:(\lambda_1,...,\lambda_{16}) \rightarrow (e^{\frac{2\pi i p_1}{N}} \lambda_1,...,e^{\frac{2\pi i p_{16}}{N}} \lambda_{16})\
\ee

The constraints that follow from demanding an orbifold of order $N$ and modular invariance are\\
\be
\sum_{i=1}^{4} r_i = \sum_{a=1}^{16} p_a =0 \, \rm{mod} \, 2 \,{\hspace*{0.05in}} \rm{for} \, {\it{N}} \, \rm{even}\
\ee
and
\be
\sum_{i=1}^{4} r_i ^2 - \sum_{a=1}^{16} p_a ^2=0
\, \rm{mod} \, {\it{N}} \, \rm{for} \, {\it{N}} \, \rm{odd}
\ee
\be
{\hspace*{1.2in}}=0 \, \rm{mod} \, 2{\it{N}} \, \rm{for} \, {\it{N}} \, \rm{even}.\
\ee
We shall use these constraints in constructing modular invariant partition functions for the non-supersymmetric $SO(32)$ theory in the later sections.\\

\section{${\mathcal{C}}/{\mathcal{Z}}_N$ flows}

  In this section we will consider heterotic string theory on ${\mathcal{R}}^{7,1} \times {\mathcal{C}}/{\mathcal{Z}}_N$ [5]. We shall start with the diagonal modular invariant non-supersymmetric $SO(32)$ theory. We shall construct various modular invariant partition functions and compute $g_{cl}$. Then we shall consider RG flows in these conformal field theories.\\
 From the constraints stated above, we see that only odd integral $N$ is allowed. This is because if $N$ is even then $r_1$ has to be even too ($r_2$=$r_3$=$r_4$=0) and so $r_1$ and $N$ are not relatively prime and we get an orbifold of order lower then $N$. Hence in the following discussion $N$ is always odd.\\
Now we construct various modular invariant partition functions for the non supersymmetric $SO(32)$ theory. We shall see that $g_{cl}$ is uniquely determined by the order of the orbifold and does not depend on the action of the orbifold group on the gauge degrees of freedom.\\
 First we consider the action of the orbifold group where $r_1 =  p_1 =1$ and the other $r_i$ and $p_a$ are all $0$. This gives rise to a (2,2) worldsheet superconformal field theory with the standard embedding of the orbifold group action into the $SO(32)$ gauge group. The modular invariant partition function is [14,15]\\
\be
Z = \frac{1}{2N} \frac{1}{\vert \eta^2 {\overline \eta}^2 {\rm{Im}} \tau \vert^3  }\sum_{k,l=0}^{N-1} \sum_{\alpha, \beta=0, \frac{1}{2}} {\frac{\zeta_{\alpha \beta}(k,l)}{\vert \theta {\frac{1}{2}+ \frac{k}{N} \brack \frac{1}{2}+ \frac{l}{N}} \vert ^2 \eta^3 {\overline \eta}^{15}}} \theta^3 {\alpha \brack \beta} \theta {\alpha + \frac{k}{N} \brack \beta + \frac{l}{N}}{\overline \theta} {\alpha + \frac{k}{N} \brack \beta + \frac{l}{N}}{\overline \theta}^{15} {\alpha \brack \beta}\
\ee
\begin{center}
where $\zeta_{00}=1, \zeta_{0 \frac{1}{2}}= \zeta_{\frac{1}{2} 0}= \zeta_{\frac{1}{2} \frac{1}{2}}=-1$
\end{center}
Next we consider the action of the orbifold group where $r_1=5, p_1=3, p_2=4 $ and the other $r_i$ and $p_a$ are all $0$. This corresponds to a (0,2) SCFT with a non-standard embedding of the orbifold action in the $SO(32)$ gauge group. The partition function is\\
\be
Z= \frac{1}{2N} \frac{1}{\vert \eta^2 {\overline \eta}^2 {\rm{Im}} \tau \vert^3} \sum_{k,l=0}^{N-1} \sum_{\alpha, \beta=0, \frac{1}{2}}{\frac{\zeta_{\alpha \beta}(k,l)}{\vert \theta {\frac{1}{2} + \frac{5k}{N} \brack \frac{1}{2} + \frac{5l}{N}} \vert ^2 \eta^3 {\overline \eta}^{15}}} \theta^3 {\alpha \brack \beta} \theta {\alpha + \frac{5k}{N} \brack \beta + \frac{5l}{N}}
\ee
\begin{center}
$$\times {\overline \theta} {\alpha + \frac{3k}{N} \brack \beta + \frac{3l}{N}}
{\overline \theta} {\alpha + \frac{4k}{N} \brack \beta + \frac{4l}{N}} {\overline \theta}^{14} {\alpha \brack \beta}$$
where $\zeta_{00}=-\zeta_{\frac{1}{2} 0}=1,  \zeta_{0 \frac{1}{2}}= \zeta_{\frac{1}{2} \frac{1}{2}}= -e^{2\pi i \frac{k}{N}}$
\end{center}

We consider one more case where $r_1=13, p_1=12, p_2=3, p_3=4$ and the other $r_i$ and $p_a$ are all 0. This too corresponds to a (0,2) theory with a nonstandard embedding of the orbifold action in the $SO(32)$ gauge group. The partition function is\\
\be
Z=\frac{1}{2N} \frac{1}{\vert \eta^2 {\overline \eta}^2 {\rm{Im}} \tau \vert^3} \sum_{k,l=0}^{N-1} \sum_{\alpha, \beta=0, \frac{1}{2}}\frac{\zeta_{\alpha \beta}(k,l)}{\vert \theta {\frac{1}{2} + \frac{13k}{N} \brack \frac{1}{2} + \frac {13l}{N}} \vert^2 \eta^3 {\overline \eta}^{15}} \theta^3 {\alpha \brack \beta} \theta {\alpha + \frac{13k}{N} \brack \beta + \frac{13l}{N}}
\ee
\begin{center}
$$\times {\overline \theta} {\alpha + \frac{12k}{N} \brack \beta + \frac{12l}{N}} {\overline \theta} {\alpha + \frac{3k}{N}\brack \beta + \frac{3l}{N}} {\overline \theta} {\alpha + \frac{4k}{N}\brack \beta + \frac{4l}{N}} {\overline \theta}^{13} {\alpha \brack \beta}$$\\
where $\zeta_{00}=-\zeta_{\frac{1}{2} 0}=1, \zeta_{0 \frac{1}{2}}=\zeta_{\frac{1}{2} \frac{1}{2}}=-e^{6\pi i \frac{k}{N}}$
\end{center}
The partition functions we have considered so far all satisfy ${r_1}^2 - \sum_{a=1}^{16} {p_a}^2=0$. Next we consider partition functions for specific choices of $N$ for the (0,2) theories where  ${r_1}^2 - \sum_{a=1}^{16} {p_a}^2=N^2$ . We shall see their importance later when we discuss RG flows. First we consider the case where $r_1=1, p_1 =p_2 =5, N=7$ and the other $r_i$ and $p_a$ are all 0. The partition function is\\
\be
Z=\frac{1}{14} \frac{1}{\vert \eta^2 {\overline \eta}^2 {\rm{Im}} \tau \vert^3} \sum_{k,l=0}^{6} \sum_{\alpha, \beta=0, \frac{1}{2}}\frac{\zeta_{\alpha \beta}(k,l)}{\vert \theta {\frac{1}{2} + \frac{k}{7} \brack \frac{1}{2} + \frac {l}{7}} \vert^2 \eta^3 {\overline \eta}^{15}} \theta^3 {\alpha \brack \beta} \theta {\alpha + \frac{k}{7} \brack \beta + \frac{l}{7}}
\ee
\begin{center}
$$\times {\overline \theta}^{2} {\alpha + \frac{5k}{7} \brack \beta + \frac{5l}{7}}  {\overline \theta}^{14} {\alpha \brack \beta}$$\\
where $\zeta_{00}=1, \zeta_{\frac{1}{2} 0}= -e^{\pi i l^2}, \zeta_{0 \frac{1}{2}}= -e^{\pi i k^2 + 9\pi i \frac{k}{7}}, \zeta_{\frac{1}{2} \frac{1}{2}}= -e^{\pi i (k^2 + l^2) + 9\pi i \frac{k}{7}}$
\end{center}
 Next we consider the case where $r_1=3, p_1=4, p_2=p_3 =3, N=5$ and the other $r_i$ and $p_a$ are all 0. The partition function is\\
\be
Z=\frac{1}{10} \frac{1}{\vert \eta^2 {\overline \eta}^2 {\rm{Im}} \tau \vert^3} \sum_{k,l=0}^{4} \sum_{\alpha, \beta=0, \frac{1}{2}}\frac{\zeta_{\alpha \beta}(k,l)}{\vert \theta {\frac{1}{2} + \frac{3k}{5} \brack \frac{1}{2} + \frac {3l}{5}} \vert^2 \eta^3 {\overline \eta}^{15}} \theta^3 {\alpha \brack \beta} \theta {\alpha + \frac{3k}{5} \brack \beta + \frac{3l}{5}}
\ee
\begin{center}
$$\times {\overline \theta}^{2} {\alpha + \frac{3k}{5} \brack \beta + \frac{3l}{5}} {\overline \theta} {\alpha + \frac{4k}{5} \brack \beta + \frac{4l}{5}} {\overline \theta}^{13} {\alpha \brack \beta}$$\\
where $\zeta_{00}=1, \zeta_{\frac{1}{2} 0}= -e^{\pi i l^2}, \zeta_{0 \frac{1}{2}}= -e^{\pi i k^2 + 7\pi i \frac{k}{5}}, \zeta_{\frac{1}{2} \frac{1}{2}}= -e^{\pi i (k^2 + l^2) + 7\pi i \frac{k}{5}}$
\end{center}
Finally we consider the case where $r_1=p_4, p_1 =3,p_2 =4,p_3=12, N=13$ and the other $r_i$ and $p_a$ are all 0. The partition function is\\
\be
Z=\frac{1}{26} \frac{1}{\vert \eta^2 {\overline \eta}^2 {\rm{Im}} \tau \vert^3} \sum_{k,l=0}^{12} \sum_{\alpha, \beta=0, \frac{1}{2}}\frac{\zeta_{\alpha \beta}(k,l)}{\vert \theta {\frac{1}{2} + \frac{r_1 k}{13} \brack \frac{1}{2} + \frac {r_1 l}{13}} \vert^2 \eta^3 {\overline \eta}^{15}} \theta^3 {\alpha \brack \beta} \theta {\alpha + \frac{r_1 k}{13} \brack \beta + \frac{r_1 l}{13}}
\ee
\begin{center}
$$\times  {\overline \theta} {\alpha + \frac{r_1 k}{13} \brack \beta + \frac{r_1 l}{13}} {\overline \theta} {\alpha + \frac{3k}{13} \brack \beta + \frac{3l}{13}} {\overline \theta} {\alpha + \frac{4k}{13} \brack \beta + \frac{4l}{13}} {\overline \theta} {\alpha + \frac{12k}{13} \brack \beta + \frac{12l}{13}} {\overline \theta}^{12} {\alpha \brack \beta}$$\\
where $\zeta_{00}=1, \zeta_{\frac{1}{2} 0}= -e^{\pi i l^2}, \zeta_{0 \frac{1}{2}}= -e^{\pi i k^2 + 19\pi i \frac{k}{13}}, \zeta_{\frac{1}{2} \frac{1}{2}}= -e^{\pi i (k^2 + l^2) + 19\pi i \frac{k}{13}}$
\end{center}
We note that there are non-trivial phases associated with the various spin structures which can be obtained by demanding modular invariance of the partition function. Following HKMM, we define $g_{cl}$ keeping only the bosonic states [16]. The formula for $g_{cl}$ is [2]
\be
Z_{\rm{tw}} ({\rm{Im}} \tau \rightarrow 0)\sim g_{cl} e^{\pi (c+ \tilde{c}) / (12 \, {\rm{Im}} \tau)}
\ee
where `tw' means that we do not include the $k=l=0$ contribution. Here, $c$ and $\tilde{c}$ refer to the effective central charges for the left and right-moving CFT respectively, and so $c=24$ and $\tilde{c} =12$ for the heterotic string. Using (17) and absorbing an overall factor of (Im${\tau)}^3$ in the definition of $g_{cl}$, we get that\\
\be
  g_{cl}= \frac{1}{2N} \sum_{k=1}^{N-1} \frac{1}{(2 {\rm{sin}} \frac{\pi k}{N} )^2}= \frac{1}{24}(N-\frac{1}{N})\
\ee
We see that the gauge degrees of freedom do not contribute to the calculation of $g_{cl}$ which is a measure of the fixed point degeneracy of the orbifold. This happens because keeping only the bosonic states in the definition of $g_{cl}$, the only contributions are from the untwisted ($\alpha = \beta =0$) and twisted ($\alpha=0, \beta = {\frac{1}{2}}$) NS sectors from the right moving CFT. In the limit Im$\tau \rightarrow 0$ , only the untwisted sector contributes and the twisted sector contribution is exponentially damped. Also $g_{cl}$ is the same for the various cases considered which is easy to see as the terms in the various summations are just permutations of each other. Note that the value of $g_{cl}$ is half of that obtained by HKMM for the type 0 theory. This is because in the HKMM analysis for the type 0 theory the contributions came from the $\alpha= \beta = 0$ and $\alpha= {\frac{1}{2}}, \beta = 0$ sectors as they are both bosonic states. However in our case the contribution comes only from the $\alpha= \beta = 0$ sector and the $\alpha= {\frac{1}{2}}, \beta =0$ sector is excluded by definition of $g_{cl}$ as that is a spacetime fermionic state. It is obvious that for any other orbifold group action consistent with modular invariance, the same value of $g_{cl}$ is obtained and we obtain the general result\\
\be
  g_{cl}(\rm{nonsusy} \, {\it{SO}}(32))=\frac{1}{24}({\it{N}}-\frac{1}{\it{N}})\
\ee
 In the HKMM paper, they had started with type 0 theory on ${\mathcal{R}}^{7,1} \times {\mathcal{C}}/{\mathcal{Z}}_N$. Unlike the heterotic case, there were no restrictions on $N$. Now let us study the RG flows associated with these ${\mathcal{C}}/{\mathcal{Z}}_N$ orbifolds. We first consider flows in theories with (2,2) superconformal invariance. Working in the RNS formalism, we construct the chiral superfield \\
\be
 \Phi = Z_1 + \overline{\theta} {\overline{\psi}}_1 + \theta \lambda_1\
\ee
 The orbifold action on the worldsheet fields is \\
\be
\Phi \rightarrow \omega^j \Phi\
\ee
where $\omega = e^{\frac{2\pi i r_1}{N}}$ and $j$=1,...,$N$-1 corresponding to the various twisted sectors of the theory. We shall refer to $r_1$ as $r$ from now on. For convenience we bosonize the fermions\\
\be
{\overline\psi}_i=e^{i {\overline H}_i}, \lambda_a=e^{i H_a}\
\ee
There are localized tachyons in each of the ($N$-1) twisted (NS-,A-) sectors. The corresponding vertex operator in the (0,-1) picture in the j-th twisted sector is [17]\\
\be
 {\mathcal{V}}_{j} = e^{-\overline{\phi}} \sigma_{\frac{j}{N} } e^{{\frac{ij}{N} } (H_1 - {\overline H}_1)} \lambda_a\
\ee
 where $\sigma_{ \frac{j}{N}  }$ is the bosonic twist operator. Here $a$=2,...,16 and $a$=1 is ruled out as that does not satisfy the level matching constraint $L_0= {\overline L}_0$. Hence, these vertex operators correspond to states in the physical Hilbert space. They give rise to tachyonic states with masses\\
\be
\frac{\alpha'}{4} {M_j}^2= -\frac{1}{2}(1- \frac{j}{N} )\
\ee
The R-charge generator corresponding to the (2,2) superconformal theory is given by $J={\lambda_1}^* \lambda_1 =i\partial H_1 $, and similarly for the right moving CFT. We consider the operators corresponding to the twists in the orbifold theory
\be
 X_{j } = \sigma_{ \frac{j}{N}} e^{ {\frac{ij}{N} } (H_1 - {\overline H}_1)} \
\ee
They have conformal dimension $ \frac{j}{2N}$ which is half their R-charge. Hence, they are elements of the (2,2) chiral ring [18]. The unperturbed chiral ring has generators
\be
 X \equiv X_1 = \sigma_{ \frac{1}{N} } e^{{\frac{i}{N} } (H_1 - {\overline H}_1)} \
\ee
and
\be
Y = \frac{1}{V} e^{i(H_1 - {\overline{H}_1})}
\ee
satisfying the chiral ring constraint $X^N =Y$. Both sides of the equation have R charge 1. Here $V$ is the volume of the noncompact orbifold, used as a regulator. We now add relevent tachyonic perturbations to the theory that preserve ${\cal{N}}$=2 superconformal invariance
\be
\delta {\cal{L}} = \lambda_j \int d^2 \theta X^j \lambda_a + c.c.
\ee
for $j$=1,...,$N-1$ and $a$=2,...,16 and find that the chiral ring constraint is undeformed! This is because the only contribution to the $X^N = Y$ constraint is the $O(\lambda_j)$ contribution by R charge conservation ($\lambda_j X^j \lambda_a$ has R charge 1) which vanishes as it involves the one point function of $\lambda_a$. Hence the chiral ring constraint is the same in the UV and the IR and suggests that there is no RG flow. Physically this happens because the (NS-,A-) tachyon is not a gauge singlet. It would be interesting to check this conclusion by other means. \\
So let us construct gauge singlet tachyonic states to study non trivial RG flows. There are such states in the (0,2) theories. In the case of the (0,2) theories the chiral superfield is given by\\
\begin{center}
$\Phi= Z_1 + \overline{\theta} \overline{\psi_1}$\\
\end{center}
The elements of the chiral ring corresponding to the various twisted sector excitations are given by
\be
X_{j}= \sigma_{ \frac{j}{N} } e^{- \frac{ij}{N} {\overline{H}}_1} \
\ee
The unperturbed chiral ring has  generators
\be
 X \equiv X_1 = \sigma_{ \frac{1}{N} } e^{-{\frac{i}{N} } {\overline H}_1} \
\ee
and
\be
Y = \frac{1}{V}  e^{-i {\overline{H}_1}} \
\ee
satisfying the chiral ring constraint $X^N =Y$. We want to perturb the UV theory by tachyonic operators that correspond to various twisted vacuua for the right and left moving CFT. The vertex operators are
\be
 {\mathcal{V}}_{j} = e^{-\overline{\phi}} X^j e^{{\frac{i}{N} } \sum_{a=1}^{16} p_a H_a} \
\ee
These are tachyonic states with masses as in (24). We now impose the GSO projection constraint for the non supersymmetric $SO(32)$ theory. The action of $(-)^{F_R}$ is given by ${\overline{H}}_1 \rightarrow {\overline{H}}_1 +N \pi$ and ${\overline{H}}_i \rightarrow {\overline{H}}_i +\pi$ for $i=2,3,4$. Here, $F_R$ is the worldsheet fermion number for the right moving CFT. The action of $(-)^{F_L}$ is given by $H_a \rightarrow H_a + N\pi$, $H_b \rightarrow H_b+ \pi$ where $a(b)=1,...,16$ refers to the gauge degrees of freedom on which the orbifold group has nontrivial (trivial) action. Here, $F_L$ is the worldsheet fermion number for the left moving CFT. Thus projecting onto the space of states satisfying $(-)^{F_R + F_L} = 1$, we get the constraint from GSO projection
\begin{center}
$j - \sum_{a=1}^{16} p_a=$ odd integer
\end{center}
Level matching constraint gives
\be
\sum_{a=1}^{16} {p_a}^2 = N^2 + j^2
\ee
We note that the level matching constraint automatically satisfies the constraint from modular invariance in (9). We now study various cases where we perturb by these tachyonic excitations and construct RG flows. Note that $p_1 \neq 0, p_2=...=p_{16}=0$ is not allowed by level matching which yields ${p_1}^2 > N^2$ whereas we have $0 < |{p_1}| < N$.\\

Next consider the case where only $(p_1, p_2) \neq 0$. As a specific example, consider the case where $p_1 = p_2 = 5, j=1, N=7$ for which the partition function has been constructed in (14). The tachyonic vertex operator is
 \be
 {\mathcal{V}}_{1} = e^{-\overline{\phi}} X e^{{\frac{5i}{7} }  (H_1 + H_2)} \
\ee
and we perturb by
\be
\delta {\mathcal{L}}= \lambda_1 \int d  {\overline{\theta}} X e^{{\frac{5i}{7} }  (H_1 + H_2)} + c.c.\
\ee
in presence of which the undeformed chiral ring constraint $X^7 = Y$ gets deformed to $X^7 + \lambda_1 X =Y$ which in the extreme IR yields $\lambda_1 X= Y$. Hence this tachyonic perturbation resolves the orbifold singularity completely and results in flat spacetime in the IR. The twisted left moving incoming vacuum gets projected onto the oppositely twisted left moving outgoing vacuum so that the total twists cancel in computing the $O(\lambda)$ correlation function.\\

Next we consider an example where only $(p_1,p_2,p_3) \neq 0$. We take $p_1=4, p_2=p_3=3, j=3,N=5$ for which the partition function has been constructed in (15). In this case the undeformed chiral ring constraint $X^5 =Y$ in presence of the perturbation gets deformed to $X^5 + \lambda_3 X^3 =Y$ hence resulting in a partial resolution of the singularity. \\

Finally we consider the case where only $(p_1,p_2,p_3,p_4) \neq 0$. Consider the case where $N=13$ corresponding to the undeformed ring relation $X^{13} =Y$ and we perturb by
\be
\delta {\mathcal{L}} = a \lambda_5 \int d \theta X^5 e^{ \frac{i}{13} (3 H_1 + 4 H_2 + 12 H_3 + 5 H_4) } + b \lambda_7 \int d \theta X^7 e^{\frac{i}{13} (3 H_1 + 4 H_2 + 12 H_3 + 7 H_4)}+c.c.
\ee
which leads to the deformed ring relation $X^{13} + a \lambda_5 X^5 + b \lambda_7 X^7 =Y $. The partition function for this orbifold action has been constructed in (16). Thus generically in the presence of relevent perturbations the chiral ring gets deformed to
\be
{(X- x_1)}^{N_1} {(X- x_2)}^{N_2}...{(X- x_k)}^{N_k}= Y\
\ee
leading to the RG flows
\be
 {\mathcal{C}}/{\mathcal{Z}}_N \rightarrow {\mathcal{C}}/{\mathcal{Z}}_{N_1} + ... +{\mathcal{C}}/{\mathcal{Z}}_{N_k}\
\ee
where $N= \sum_{i=1}^{k} N_i$ along which $g_{cl}$ decreases monotonically. Here all $N_i$ are odd integers to have a consistent interpretation of the extreme IR theory as the orbifold CFT of the heterotic string theory.\\

Next we would like to study the RG flow in supersymmetric $SO(32)$ and $E_8 \times E_8$ heterotic string theories. We start with the non-supersymmetric $SO(32)$ theory and perform a twist by ${(-)}^{F_R}$ to get the supersymmetric $SO(32)$ theory and a further twist by ${(-)}^{F_1}$ to get the $E_8 \times E_8$ theory. Here, $F_1$ is the world sheet fermion number which anticommutes with $\lambda_1,...,\lambda_8$ in the left-moving CFT.\\
First let us consider the supersymmetric $SO(32)$ theory in detail. We start with nonsupersymmetric $SO(32)$ and gauge the $Z_2$ symmetry ${(-)}^{F_R}$. The action of ${(-)}^{F_R}$ on ${\overline{H}}_1$ is ${\overline{H}}_1 \rightarrow {\overline{H}}_1 +N \pi$ as stated before. Consequently, states where $j$ is an odd integer survive the GSO projection. This leads to a deformed chiral ring relation in the IR
\be
{X}^{N'} {(X- x_1)}^{N_1} {(X+ x_1)}^{N_1}...{(X- x_k)}^{N_k}{(X+ x_k)}^{N_k} = Y\
\ee
where $N= N' +2 \sum_{i=1}^{k} N_i$ and $N, N', N_i$ are all odd integers. So the vacuum at $X=0$ is a supersymmetric $SO(32)$ vacuum while in the remaining vacuua at $X=x_i$, the chiral GSO is spontaneously broken and one has nonsupersymmetric SO(32) theories in these vacuua. So the RG flow leads to the following decay of spacetime
\be
\rm{susy} \, {\it{SO}}(32) \rightarrow \rm{susy} \, {\it{SO}}(32) + \rm{nonsusy} \, {\it{SO}}(32)s\
\ee
This is again analogous to the HKMM RG flow, where one starts with type II theory in the UV and ends with a type II vacuum and many type 0 vacuua in the extreme IR. To check the $g_{cl}$ conjecture, following HKMM we would like to define $g_{cl}$ in the supersymmetric theory keeping only spacetime bosonic states. Consequently the contributions to $g_{cl}$ in the limit Im$\tau \rightarrow 0$ in (17) for the diagonal modular invariant non-supersymmetric $SO(32)$ theory comes from the $\alpha=0, \beta=0$ sector only as explained after (18). The GSO projection contributes a factor of $\frac{1}{2}$ in the partition function to give an overall factor of $\frac{1}{2}$. For the supersymmetric $SO(32)$ theory there is a further twist by $(-)^{F_R}$ and so the partition function has a factor of $\frac{1}{4}$. However in this case there are two contributions to $g_{cl}$: (i) $\alpha= \beta=0; \gamma =\delta =0$ and (ii) $\alpha= \beta=0; \gamma ={\frac{1}{2}} ,\delta =0$ where $\alpha, \beta$ $(\gamma ,\delta)$ refer to the spin structures corresponding to the right (left) moving CFT and $\alpha, \gamma$ refer to the upper index in the theta function and $\beta, \delta$ refer to the lower index. So the total factor is again $\frac{1}{2}$ and we find that $g_{cl}$ for non-supersymmetric $SO(32)$ is equal to $g_{cl}$ for supersymmetric $SO(32)$. In this case \\
\be
g_{cl}^{UV} = \frac{1}{24} (N-\frac{1}{N})\
\ee
\be
g_{cl}^{IR} = \frac{1}{24} (N'-\frac{1}{N'}) + \frac{1}{24} \sum_{i=1}^{k} (N_i -\frac{1}{N_i})\
\ee
In the calculation of $g_{cl}$, we consider the vacuum either at $X=x_i$ or at $X=-x_i$ but not both, as they are related by the broken chiral GSO. Again we see that $g_{cl}$ decreases along the flow and hence the conjecture is satisfied.\\

Finally let us consider RG flows in the $E_8 \times E_8$ heterotic string theory. This is obtained by gauging the $Z_2$ action ${(-)}^{F_1}$ in the supersymmetric $SO(32)$ theory. The action of the  ${(-)}^{F_1}$ GSO on $\lambda_a$ $(a=1,...,8)$ is \\
\be
  H_b \rightarrow  H_b + N\pi, H_c \rightarrow  H_c + \pi \
\ee
where $b(c)=1,...,8$ refer to the gauge degrees of freedom on which the orbifold group has nontrivial (trivial) action. Under the action of ${(-)}^{F_1}$, by looking at the vertex operator (32) corresponding to the tachyonic states we see that only states with $\sum_{a=1}^{8} p_a=$ even integer survive the GSO projection. Also we have seen that under the ${(-)}^{F_R}$ GSO only states with $j=$ odd integer survive. So in the $E_8 \times E_8$ theory, only states where $j$ is an odd integer and $\sum_{a=1}^{8} p_a$ is an even integer survive the GSO projection while the remaining states are projected out of the spectrum. Hence, the RG flow proceeds as before leading to\\
\be
E_8 \times E_8 \rightarrow E_8 \times E_8 + \rm{nonsusy} \, {\it{SO}}(32)s\
\ee
Again we define $g_{cl}$ for the $E_8 \times E_8$ theory keeping only spacetime bosonic states in the spectrum. In this case, there is a further twist by $(-)^{F_1}$ and so the partition function has a factor of $\frac{1}{8}$. However there are now 4 contributions to $g_{cl}$ in the limit Im$\tau \rightarrow 0$: (i) $\alpha= \beta=0; \gamma = \delta =0; \omega= \rho= 0,$ (ii) $\alpha= \beta=0; \gamma = \delta =0; \omega={\frac{1}{2}}; \rho= 0,$ (iii) $\alpha= \beta=0; \gamma ={\frac{1}{2}}; \delta =0; \omega= \rho= 0,$ and (iv) $\alpha= \beta=0; \gamma ={\frac{1}{2}}; \delta =0; \omega={\frac{1}{2}}; \rho= 0$ where $\alpha, \beta$ refer to the spin structures corresponding to the right moving CFT and $\gamma, \delta$ $(\omega, \rho)$ refer to the spin structures corresponding to the left moving CFT involving $\lambda_1,...,\lambda_8$ $(\lambda_9,...,\lambda_{16})$ and $\alpha, \gamma, \omega$ refer to the upper index in the theta function and $\beta, \delta, \rho$ refer to the lower index. So the total factor is $\frac{1}{2}$ and $g_{cl}$ for the  $E_8 \times E_8$ theory is equal to $g_{cl}$ for the non-supersymmetric $SO(32)$ theory. This leads to
\be
g_{cl}^{UV}= \frac{1}{24} (N - \frac{1}{N})\
\ee
\be
g_{cl}^{IR} = \frac{1}{24} (N' - \frac{1}{N'}) + \frac{1}{24} \sum_{i=1}^{k} (N_i - \frac{1}{N_i})\
\ee
where $N = N' + 2\sum_{i=1}^{k} N_i$. Again the $g_{cl}$ conjecture is satisfied.\\

\section{${\mathcal{C}}^2/{\mathcal{Z}}_N$ flows}
  In this section we consider ${\mathcal{C}}^2/{\mathcal{Z}}_N$ orbifolds of heterotic string theory and study the RG flows. A computation of $g_{cl}$ again shows that it decreases monotonically along RG flows. We consider actions of the orbifold group where $r_1 =1$ in order to use the Hirzebruch Jung theory of singularity resolutions [2,19,20,21]. Demanding that we have an orbifold of order $N$ , we find that $r_2$ has to be an odd integer. We shall refer to $r_2$ as $r$ from now on and denote the orbifold as ${\mathcal{C}}^2/{\mathcal{Z}}_{N(r)}$ as in [1].  We note that for these theories $N$ can be either even or odd. Specifically once again we are interested in (2,2) and (0,2) theories.\\
We consider modular invariant partition functions for the non-supersymmetric $SO$(32) theory. We consider the case where $r_1= p_1=1, r_2= p_2=\pm1$ and the other $r_i$ and $p_a$ are all 0. This corresponds to a (2,2) theory with the standard embedding of the orbifold action in the $SO$(32) gauge group. The partition function is\\
\be
Z= \frac{1}{2N} \frac{1}{\vert \eta^2 {\overline \eta}^2 {\rm{Im}} \tau \vert^2} \sum_{k,l=0}^{N-1} \sum_{\alpha, \beta=0, \frac{1}{2}} \frac{\zeta_{\alpha \beta}(k,l)}{\eta^2 {\overline \eta}^{14} \vert \theta {\frac{1}{2} + \frac{k}{N} \brack \frac{1}{2} + \frac{l}{N}} \theta {\frac{1}{2} \pm \frac{k}{N} \brack \frac{1}{2} \pm \frac{l}{N}} \vert^2} \theta^2 {\alpha \brack \beta} \theta {\alpha + \frac{k}{N} \brack \beta + \frac{l}{N}}\theta {\alpha \pm \frac{k}{N} \brack \beta \pm \frac{l}{N}}\
\ee
\begin{center}
$$\times {\overline \theta} {\alpha + \frac{k}{N} \brack \beta + \frac{l}{N}} {\overline \theta} {\alpha \pm \frac{k}{N} \brack \beta \pm \frac{l}{N}} {\overline \theta}^{14} {\alpha \brack \beta}$$
\end{center}
\begin{center}
where $\zeta_{00}=-\zeta_{\frac{1}{2} 0}=-\zeta_{0 \frac{1}{2}}=-\zeta_{\frac{1}{2} \frac{1}{2}}=1$.\\
\end{center}
Keeping only the bosonic states and absorbing an overall factor of (Im${\tau)}^2$ in the definiton of $g{_{cl}}$ in (14), we get that [2]\\
\be
g_{cl}(N,\pm1)=\frac{1}{2N} \sum_{k=1}^{N-1} \frac{1}{(4{\rm{sin}}^2 \frac{\pi k}{N})^2}= \frac{(N^2 +11)(N^2 -1)}{45 \cdot 32N}\
\ee
Again the value of $g_{cl}$ is half of that obtained by HKMM for the type 0 theory for reasons explained after (18) and we note that $g_{cl}$ is a monotonically decreasing function of $N$.
Next we consider the case where $r_1=1, r_2=\pm3, p_1=p_2=1, p_3=p_4=2$ and the other $r_i$ and $p_a$ are all 0. This corresponds to a (0,2) theory with a non-standard embedding of the orbifold action in the $SO$(32) gauge group. The partition function is\\
\be
Z=\frac{1}{2N} \frac{1}{\vert \eta^2 {\overline \eta}^2 {\rm{Im}} \tau \vert^2} \sum_{k,l=0}^{N-1} \sum_{\alpha, \beta=0, \frac{1}{2}} \frac{\zeta_{\alpha \beta}(k,l)}{\eta^2 {\overline \eta}^{14} \vert \theta {\frac{1}{2} + \frac{k}{N} \brack \frac{1}{2} + \frac{l}{N}}\theta {\frac{1}{2} \pm \frac{3k}{N} \brack \frac{1}{2} \pm \frac{3l}{N}}\vert^2} \theta^2 {\alpha \brack \beta}\theta {\alpha + \frac{k}{N} \brack \beta + \frac{l}{N}} \theta{\alpha \pm \frac{3k}{N} \brack \beta \pm \frac{3l}{N}}\
\ee
\begin{center}
$$\times {\overline \theta}^2 {\alpha + \frac{k}{N} \brack \beta + \frac{l}{N}} {\overline \theta}^2 {\alpha + \frac{2k}{N} \brack \beta + \frac{2l}{N}}  {\overline \theta}^{12}{\alpha \brack \beta}$$
\end{center}
\begin{center}
where $\zeta_{00}=-\zeta_{\frac{1}{2} 0}=1, \zeta_{0 \frac{1}{2}}= \zeta_{\frac{1}{2} \frac{1}{2}}=-e^{2\pi i \frac{k}{N}}$ if $r_2=3$\\
and $\zeta_{00}=-\zeta_{\frac{1}{2} 0}=1, \zeta_{0 \frac{1}{2}}= \zeta_{\frac{1}{2} \frac{1}{2}}=-e^{8\pi i \frac{k}{N}}$ if $r_2=-3$\
\end{center}
We get the following expression for $g_{cl}$\\
\be
g_{cl}(N,\pm3)=\frac{1}{2N} \sum_{k=1}^{N-1}\frac{1}{(4{\rm{sin}} \frac{\pi k}{N} {\rm{sin}} \frac{3\pi k}{N})^2}\
\ee
This sum can be exactly evaluated [2] to give
\be
  g_{cl}(N,\pm3)= \frac{(N^4 + 210 N^2 - 80 N - 291)}{405 \cdot 32 N} \,{\hspace*{0.1in}} {\rm{for}} \, N = 2 \, {\rm mod}\,  3
\ee
\be
 {\hspace*{0.8in}}=\frac{(N^4 + 210 N^2 + 80 N - 291)}{405  \cdot 32 N} \, {\hspace*{0.1in}}{\rm{for}} \, N = 1 \, {\rm mod} \, 3
\ee
Again the value of $g_{cl}$ is half of that obtained by HKMM for the type 0 theory for reasons explained after (18). Also we note that $g_{cl}$ is a monotonically decreasing function of $N$.\\
We now construct RG flows corresponding to the various heterotic string theories. Again the results are similar to the HKMM analysis and we shall be brief.\\
Corresponding to the ${\mathcal{C}}^2/{\mathcal{Z}}_{N(r)}$ orbifold, the elements of the $\mathcal{N}$=2 chiral ring corresponding to the various twisted sectors are given by
\be
 X_j = {X_j}^{(1)} {X_{N \{ \frac{rj}{N} \} }}^{(2)}
\ee where 1 and 2 refer to $Z^1$ and $Z^2$ and for the (2,2) theories $a=3,...,16$ as explained below (55). The chiral GSO ${(-)}^{F_R}$ has the action
\be
 {\overline{H}}_1 \rightarrow {\overline{H}}_1 +r \pi,{\hspace*{0.2in}} {\overline{H}}_2 \rightarrow {\overline{H}}_2 - \pi
\ee
Hence states where [$\frac{rj}{N}$] is an even integer are projected out while states where [$\frac{rj}{N}$] is an odd integer are kept in the spectrum. We consider ${\mathcal{C}}^2/{\mathcal{Z}}_{N(\pm1)}$ orbifolds and construct RG flows to check the $g_{cl}$ conjecture.
  We consider the (2,2) non-supersymmetric $SO$(32) theory on ${\mathcal{C}}^2/{\mathcal{Z}}_{N(1)}$ orbifold. In the (0,-1) picture the tachyonic vertex operator in the $j$-th twisted (NS-,A-) sector is given by
\be
{\mathcal{V}}_j=e^{-{\overline{\phi}}} {X_{j}}^{(1)} {X_{j}}^{(2)} \lambda_a
\ee
Here $a$=3,...,16 ($a$=1,2 does not satisfy level matching) and the unperturbed chiral ring satisfies the relation
\be
X^N = Y^{(1)} Y^{(2)}
\ee
where
where $X \equiv X_1$. $X_j$ has R charge $\frac{2j}{N}$ and conformal dimension $\frac{j}{N}$. Hence perturbations where $j < \frac{N}{2}$ are relevent and those with $j= \frac{N}{2}$ (for $N$ even) are marginal. They lead to the deformed chiral ring relation
\be
X^N  +  \sum_{j+k < N}  c_{jk} \lambda_j \lambda_k X^{j+k}= Y^{(1)} Y^{(2)}
\ee
The $O(\lambda_j)$ corrections vanish by R-charge conservation as all the terms have R-charge 2. The $O({\lambda_j}^2)$ contribution is nonvanishing as it involves the gauge singlet two point function of $\lambda_a$ and ${\lambda_a}^*$. Hence in the presence of relevent perturbations, the UV theory flows to a ${\mathcal{C}}^2/{\mathcal{Z}}_{M(1)}$ theory in the IR where $M<N$. Hence the $g_{cl}$ conjecture is satisfied. For marginal perturbations there is no flow. Under the chiral GSO ${(-)}^{F_R}$ the entire chiral ring is projected out and so there are no RG flows in the supersymmetric $SO$(32) and $E_8 \times E_8$ string theories. For the (0,2) theories, the RG flows remain the same and the level matching constraint restricts the choice of $p_a$s.\\
For the ${\mathcal{C}}^2/{\mathcal{Z}}_{N(-1)}$ flows again we start with the non-supersymmetric $SO$(32) theory. There are no tachyons in the twisted (NS-,A-) sectors as they are massles and all the perturbations are marginal. However, perturbing the theory by marginal operators and sending the moduli corresponding to the various twisted sectors to infinity, the orbifold theory can be decoupled to two orbifold theories
\be
{\mathcal{C}}^2/{\mathcal{Z}}_{N(-1)} \rightarrow {\mathcal{C}}^2/{\mathcal{Z}}_{j(-1)} + {\mathcal{C}}^2/{\mathcal{Z}}_{N-j(-1)}
\ee
along which $g_{cl}$ decreases. In this case the modulus corresponding to the $j$-th twisted sector has been sent to infinity. We now consider the (2,2) theories. Under the chiral GSO, the entire chiral ring survives and the flow proceeds as above in the supersymmetric $SO$(32) theory. Further twisting by ${(-)}^{F_1}$ to get the $E_8 \times E_8$ theory, ${(-)}^{F_1}$ has the action
\be
H_1 \rightarrow H_1 + r \pi,  {\hspace*{0.2in}}  H_2 \rightarrow H_2 - \pi\
\ee
Considering the (NS-,A-) sector twisted vertex operators, we see that states where $a$=3,...,8 ($a$=1,2 does not satisfy level matching) survive the GSO projection  while states where $a$=9,...,16 are projected out. In the (0,2) case the RG flows remains the same and the choices of $p_a$s are restricted by the level matching constraint.\\

For the ${\mathcal{C}}^2/{\mathcal{Z}}_{N(\pm3)}$ orbifolds, the analysis is similar to the HKMM analysis with constraints on the gauge degrees of freedom as mentioned above. Also $g_{cl}$ decreases along RG flows.

\section{$AdS_3/{\mathcal{Z}}_N$ flows}
We shall consider type 0 string theory propagating on $AdS_3/{\mathcal{Z}}_N \times \mathcal{M}$ [22,9,23,24], where $\mathcal{M}$ is a seven dimensional manifold. It is known that a $\mathcal{Z_{N}}$ orbifold of $AdS_3$ [9,10] breaks spacetime supersymmetry completely and there are twisted sector tachyons in the NS sector which are localized at the orbifold fixed point. The vertex operators corresponding to the tachyonic excitations have been constructed [9]. The vertex operator corresponding to the tachyon in the $q$-th twisted sector ($q$=1,...,$N$-1) is given by
\be
{\hat{\Psi}}_{j_q, j_q, j_q} = \Psi_{j_q, j_q, j_q} e^{-i \frac{q}{N} (H + \sqrt{\frac{2}{\tilde{k}}}Y) -i \frac{q}{N} ({\overline{H}} + \sqrt{\frac{2}{\tilde{k}}}{\overline{Y}})}
\ee
where $j_q = \frac{kq}{2N}$ and $\tilde{k} =k +2$. $H$ is given by
\be
H=\sqrt{\frac{k}{2}} {\mathcal{Y}} - \sqrt{\frac{\tilde{k}}{2}} Y
\ee
where
\be
j^3 = i \sqrt{\frac{\tilde{k}}{2}} \partial Y
\ee
and
\be
J^3 = i \sqrt{\frac{k}{2}} \partial {\mathcal{Y}}
\ee
Here $j^a (J^a)$ are the bosonic (total) currents associated with the $\widehat{SL}$(2,$R$) current algebra. Also $\tilde{k}$ is the level of the supersymmetric WZW model and $k$ is the level of the bosonic WZW model. $\Psi_{j_q, j_q, j_q}$ is a parafermion corresponding to the $q$th twisted sector having conformal dimension $\frac{\tilde{k}}{2(\tilde{k} -2)} \frac{q}{N} (1- \frac{q}{N})$. ${\hat{\Psi}}_{j_q, j_q, j_q}$ has conformal dimension $\frac{q}{2N}$ and is the vertex operator corresponding to tachyonic excitations with mass
\be
\frac{\alpha '}{4} {M_q}^2=-\frac{1}{2} (1- \frac{q}{N})
\ee
The worldsheet theory has (2,2) superconformal invariance generated by [22]
\be
T= \frac{1}{k} {\hspace*{0.04in}} g_{ab} {\hspace*{0.04in}}  (j^a j^b - \psi^a \partial \psi^b)
\ee
\be
G= \frac{2}{k}  {\hspace*{0.04in}} (g_{ab} \psi^a j^b - \frac{i}{3k} \epsilon_{abc} \psi^a \psi^b \psi^c)
\ee
and
\be
J= \frac{1}{k}  {\hspace*{0.04in}} (-i \tilde{k} \partial H + 2 j^3)
\ee
where $\psi^a$ is the bottom component of the WZW supercurrent. The metric is $g^{ab}$=diag(1,1,-1) and $\epsilon^{123}$=1. We see that the conformal dimension of ${\hat{\Psi}}_{j_q, j_q, j_q}$ is half the R-charge and so they are elements of the $\mathcal{N}$=2 chiral ring. Hence one can compute $g_{cl}$ and study RG flows that preserve $\mathcal{N}$=2 superconformal invariance. The details will mimic the HKMM analysis and we shall not repeat it here.\\
The partition function for type 0 string theory on $AdS_3/{\mathcal{Z}}_N \times \mathcal{M}$ can be computed [25,24,10] and is given by
\be
Z= \frac{1}{2N} \frac{1}{\vert \eta \vert^2 {\sqrt {{\rm{Im}} \tau}} } \sum_{k,l=0}^{N-1} \sum_{\alpha, \beta=0, \frac{1}{2}} \frac{ Z_{\mathcal{M}}}{{\eta^3 {\overline \eta}^{3}} \vert \theta {\frac{1}{2} + \frac{k}{N} \brack \frac{1}{2} + \frac{l}{N}} \vert^2} \vert \theta^3 {\alpha \brack \beta} \theta {\alpha + \frac{k}{N} \brack \beta + \frac{l}{N}} \vert^2  \
\ee

Here $Z_{\mathcal{M}}$ is the partition function involving the bosonic CFT on $\mathcal{M}$. We can compute $g_{cl}$ for $AdS_3/{\mathcal{Z}}_N$ using (17) and absorbing an overall factor of $\sqrt{{\rm{Im}} \tau}$ in the definiton of $g_{cl}$ to get
\be
g_{cl} = \frac{1}{12} (N - \frac{1}{N})
\ee
Let us briefly study the deformation theory of the chiral ring to see how the decay of spacetime occurs. As mentioned earlier, the vertex operators corresponding to the various twisted sector tachyonic excitations have been constructed in [9] to which we refer the reader for details. Hence the undeformed chiral ring has two generators
\be
X \equiv {\hat{\Psi}}_{j_1, j_1, j_1} = \Psi_{j_1, j_1, j_1} e^{- \frac{i}{N} (H + \sqrt{\frac{2}{\tilde{k}}}Y) - \frac{i}{N} ({\overline{H}} + \sqrt{\frac{2}{\tilde{k}}}{\overline{Y}})}
\ee
and
\be
Y = {\frac{1}{V}} e^{-i  (H + \sqrt{\frac{2}{\tilde{k}}}Y) -i ({\overline{H}} + \sqrt{\frac{2}{\tilde{k}}}{\overline{Y}})}
\ee
where $\Psi_{j_1, j_1, j_1}= \sigma_1$ is the parafermionic unit twist operator and the underformed ring relation is $X^{N} =Y$ which in the presence of relevent perturbations get deformed to
\be
{(X- x_1)}^{N_1} {(X- x_2)}^{N_2}...{(X- x_k)}^{N_k}= Y\
\ee
leading to the RG flows
\be
 {\mathcal{C}}/{\mathcal{Z}}_N \rightarrow {\mathcal{C}}/{\mathcal{Z}}_{N_1} + ... +{\mathcal{C}}/{\mathcal{Z}}_{N_k}\
\ee
where $N= \sum_{i=1}^{k} N_i$ along which $g_{cl}$ decreases monotonically and so the $g_{cl}$ conjecture is satisfied. We note that $g_{cl}$ is exactly the same as that for type 0 theory on flat space as calculated in [2]. This is because the bosonic determinant in the denominator of the partition function which enters the computation of $g_{cl}$ is the same as that in the flat space theory. The free fermionic determinants are the same as well. Following the lines in the HKMM analysis, the same conclusion follows for the type II theory as well.
\section{Conclusion}
Following the conjecture made by HKMM, we have constructed RG flows in heterotic string theories and $\mathcal{Z_{N}}$ orbifold of $AdS_3$. We have considered a class of theories having $\mathcal{N}$=2 worldsheet superconformal invariance and calculated $g_{cl}$, which is a monotonically decreasing function along RG flows- in accordance with the conjecture made by HKMM. In the case of the heterotic string theories, the gauge degrees of freedom do not contribute to  the computation of $g_{cl}$. This is evident from the structure of the modular invariant partition functions that have been constructed. The bosonic determinants in the denominator contribute non-trivially to the computation of $g_{cl}$ while the world sheet fermionic determinants in the numerator merely contribute a numerical factor as Im$\tau \rightarrow 0$. Hence the gauge degrees of freedom which contribute as fermionic determinants in the partition function do not play a non-trivial role in the computation of $g_{cl}$.

\section*{Acknowledgement}
I am grateful to Prof. Jeffrey A. Harvey for suggesting the problem and for many enlightening discussions in the course of the project.

\end{document}